\documentclass[letterpaper, 10 pt, conference]{ieeeconf}
\IEEEoverridecommandlockouts

\usepackage{graphicx}
\usepackage{caption}
\usepackage{algpseudocode}
\usepackage{amsmath}
\usepackage{longtable}
\usepackage{adjustbox}

\usepackage{amsmath,amssymb,amsfonts,epsf,epsfig,times}
\usepackage[all]{xy}
\usepackage{graphicx,color}
\usepackage{subfigure}
\usepackage{url}
\usepackage{cite}
\usepackage{tikz}
\usepackage{epstopdf}
\usepackage{algorithm}
\usepackage[]{amssymb}

\def\QED{\mbox{\rule[0pt]{1.5ex}{1.5ex}}}

\newcommand{\OMIT}[1]{}

\makeatletter
\def\BState{\State\hskip-\ALG@thistlm}
\makeatother

\begin{document}
\title{Multi-Objective Cooperative Search of Spatially Diverse Routes in Uncertain Environments}
\author{Johnathan Votion and Yongcan Cao
\thanks{The authors are with the Department of Electrical and Computer Engineering, The University of Texas, San Antonio, TX 78249. 
}
\thanks{Corresponding Author: Yongcan Cao (yongcan.cao@utsa.edu)}
}

\markboth{}
         {}

\maketitle

\begin{abstract}
This paper focuses on developing new navigation and reconnaissance capabilities for cooperative unmanned systems in uncertain environments. The goal is to design a cooperative multi-vehicle system that can survey an unknown environment and find the most valuable route for personnel to travel. To accomplish the goal, the multi-vehicle system first explores spatially diverse routes and then selects the safest route. In particular, the proposed cooperative path planner sequentially generates a set of spatially diverse routes according to a number of factors, including travel distance, ease of travel, and uncertainty associated with the ease of travel. The planner's dependence on each of these factors is altered by a weighted score, doing so changes the criteria for determining an optimum route. To penalize the selection of same paths by different vehicles, a control gain is used to increase the cost of paths that lie near the route(s) assigned to other vehicles. By varying the control gain, the spatial diversity among routes can be accomplished. By repeatedly searching for different paths cooperatively, an optimal path can be selected that yields the most valuable route.
\end{abstract}


\IEEEpeerreviewmaketitle


\section{Introduction}

The recent capabilities of unmanned vehicles and robotic systems have increased their usefulness in military and rescue applications. For example, small UAVs are now being utilized to provide aerial surveillance in combat areas. This work introduces a cooperative path planning method that allows a team of autonomous vehicles to be instructed to conduct reconnaissance throughout an environment in order to determine a safe, low-risk route to a target location. Most unmanned vehicle systems experience degraded performance when executing these cooperative missions due to the approximation of map information and inadequete scoring algorithms used by the path planner.


In many applications, robotic systems need to take responsive actions in unknown/dynamic environments where both situational awareness and path planning need to be properly addressed. For example, the research work in~\cite{bib:CoverIntro,bib:Cover1} focused on properly assigning mobile sensors' motion in order to maximize the detection probability with a given distribution density function that indicates the probability that some event takes place in a region. The existing work in~\cite{bib:RiskLit1,bib:RiskLit2,bib:RiskLit3,bib:RiskPlan1,bib:RiskLit5,bib:Adverse1}  focuses on planning paths of robotic systems in uncertain and dynamic environments. For example, one main research topic is to find the shortest path from a starting position to a target location~\cite{bib:PathPlan10, bib:Roadmap5, bib:PathPlan4, bib:PathPlan5, bib:PathPlan6B, bib:PathPlan7, bib:PathPlan8, bib:PathPlan9, bib:Short1}, where numerous algorithms, such as A search, heuristic methods, Dijksta's algorithms, and Voronoi Diagram~\cite{bib:Voronoi1}, are used to create paths with minimum costs. 

Although the existing research provides numerous methods to derive paths for robotic systems, there has been very limited work considering the selection of diverse enough routes such that rich knowledge can be obtained to find the safest path. Current methods cannot be easily adjusted to select how diverse the paths should be under different scenarios. For example, in disaster response missions, unmanned vehicles that are used to search for the safest path should coordinate their path planning efforts to ensure that minimum overlap occurs when exploring the environment. When the environment is unknown, it becomes very challenging to plan paths for multiple vehicles cooperatively to satisfy both path optimization and path diversity simultaneously. Considering the different optimization criteria, such as travel distance, ease of travel, and uncertainty associated with the ease of travel, becomes a complex problem in determining the safest route.

The objective of the paper is to derive a uniform path planning algorithm that can address numerous criteria that are of interest to the system operator for different missions. To accomplish the objective, the robotic system need to explore spatially diverse paths for the subsequent selection of the safest route. The proposed cooperative path planner seeks to generate a set of spatially diverse routes sequentially based on numerous factors of interest, such as travel distance, ease of travel, and uncertainty associated with the ease of travel. The proposed path planner allows the users to create unique sets of planned paths by changing parameters appropriately. To enable the selection of diverse paths by different vehicles, a control gain is used to increase the cost of paths that lie near the route(s) assigned to other vehicles. By varying the control gain, the spatial diversity among routes can be accomplished. For an unknown environment, repeatedly searching spatially diverse paths can yield the most valuable route.

The remainder of this paper is organized as follows: In section II, the problem formulation is described. In section III preliminary definitions are given. In section IV, the proposed method is described. In section V, simulations are made to observe the performance of the proposed method. Finally, in section VI, the conclusion for this work is presented.

\section{Problem Formulation}
Acting as the first wave of on-site responders, a team of autonomous unmanned vehicles are instructed to find a ``good quality'' route so that human personnel can safely travel through an environment. The quality of the route is determined by its traversability (i.e.~ease of travel) and total length. The team of vehicles are tasked to explore the environment and identify (from multiple possible routes) a safe path to a target location. 

The routes assigned to the team are calculated in iterations, with each route calculated to take a substantially different path through the map than any path defined in a previous iteration. The level of similarity between calculated paths is selected via a control gain, where high gain values result in increased `dissimilarity' between routes. Searching dissimilar routes throughout the map is important because it allows the unmanned system to obtain more map information then the coverage provided by a single path. It also improves the vehicle's reaction to unforseen hazards. For example, the initial path assigned to a vehicle may be obstructed such that the vehicle is required to reroute its path. In this case the vehicle may use updated map information obtained through inter-vehicle communication to generate improved plans.

Figure 1 illustrates how a team of unmanned vehicles can perform a cooperative mission where a safe route to a target location is identified by exploring dissimilar routes (shown as the red trajectories). Initial paths for the vehicles are generated using \textit{a priori} data (e.g. satellite or high altitude imagery). Vehicles start from the blue marker, which in this example is an offshore location, and make their way along assigned paths to the target location (denoted with the black marker). The unmanned vehicles are assumed to be either aerial or amphibious. Along the route, vehicles survey the environment and update their maps accordingly. After reaching the target location, the unmanned vehicles communicate their data to an operator that can direct personnel along the safest calculated path (shown as the black trajectory).

\begin{figure}[htb!]
\centering
\includegraphics[scale=0.33]{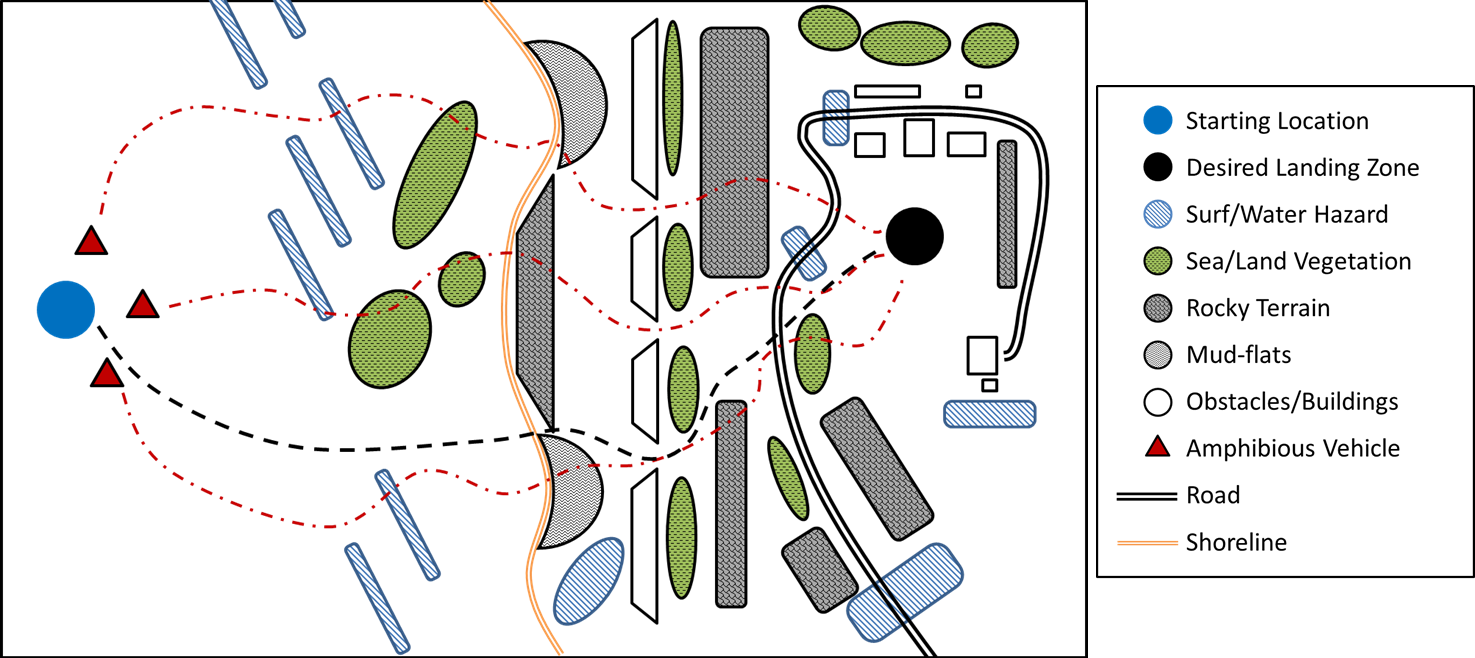}
\caption{Satellite image for mission area}
\label{fig:satt}
\end{figure}

Because the terrain is unknown, the environment poses a number of hazards to the unmanned systems. In addition the \textit{a priori} information initially used may not be consistent with changes that have recently occurred. For example, the tide may change the position of the shoreline, vegetation may have grown abundant in concentrated areas, or fallen trees may obstruct paths that were expected to be clear.

The operator may be confronted with a particular situation in which he or she must prioritize path conditions (e.g. distance of the path vs.~ease of travel of the path). To evaluate routes given prioritized conditions, the path planner uses weighted gains. These weighted gains balance the planner's dependencies on traversability, travel distance, and uncertainty of the path.

\section{PRELIMINARIES}

Let $\mathcal{A}\triangleq \{A_1,\cdots , A_{N_A}\}$ represent a set of $N_A\in \mathbb{Z}_{>0}$ vehicles in an unmanned system. Each vehicle is represented as a particle in a 2-space environment with respect to the $x$ and $y$ dimensions. Any position in the environment may be described by
\begin{equation}
r=[r_x~r_y]^T
\end{equation}
where $r\in \mathbb{R}^2$ is the position vector.

\subsection{Defining the Roadmap}
\begin{figure}[htb!]
\centering
\includegraphics[scale=0.3]{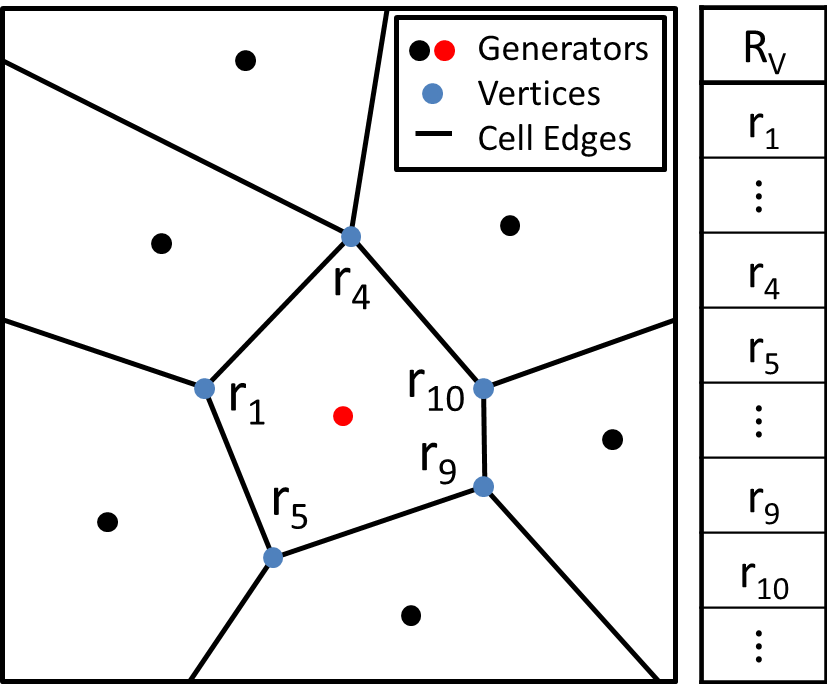}
\caption{Voronoi cell decomposition example}
\label{fig:roadmap1}
\end{figure}

As presented in \cite{bib:Votion_Diss}, a Voronoi diagram approach is used to define the roadmap. In the 2D case, an ordinary Voronoi diagram is a partitioning of a plane into regions based on the euclidean distance to points in a specific subset of the plane. The Voronoi diagram is described by a number of points called generators, and their corresponding Voronoi cells. A Voronoi diagram is defined such that any point within a Voronoi cell is closer to that cell's corresponding generator than it is to any other generator. Figure \ref{fig:roadmap1} shows a Voronoi diagram for some arbitrary placement of generators, where $R_V$ is a list of vertex positions. The original work associated with the Voronoi diagram can be found in \cite{bib:Voronoi1} and \cite{bib:Voronoi2}. A similar approach for defining generators is used in \cite{bib:Roadmap6}.

\begin{figure}[htb!]
\centering
\includegraphics[scale=0.6]{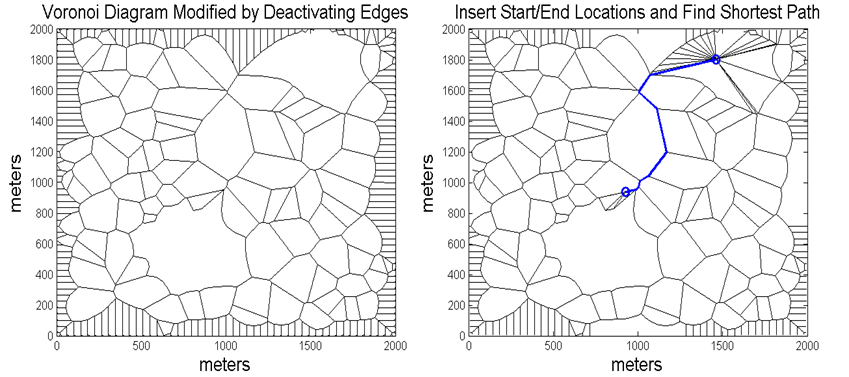}
\caption{Modified Voronoi diagram and calculated route between start/end points}
\label{fig:roadmap4}
\end{figure}

The Voronoi diagram is converted into roadmap information by treating cell edges as possible paths and vertices as waypoints. The roadmap is modified by removing superfluous edges and inserting start and end points. Figure \ref{fig:roadmap4} illustrates this process, showing (in blue) the shortest path from the inserted start and end points. The roadmap information is defined by the undirected adjacency matrix $G\in \mathbb{R}^{(N_W\times N_W)}$ and cost matrix $C\in \mathbb{R}^{(N_W\times N_W)}$, where $N_W$ is the total number of waypoints. The elements of the adjacency matrix are defined as $G_{nn} = 0$, $G_{nm,n\neq m} = 1$ if and only if there exists an edge between the vertex positions listed in the $n^{th}$and $m^{th}$ row of $R_V$, and $G_{nm,n\neq m} = 0$ otherwise. A cost value is calculated for each edge defined in $G$ and recorded into the corresponding position of $C$. Using $G$ and $C$ the path planner (e.g. \textit{A*}) can generate a ``good quality'' path (this work uses the shortest path algorithm given in \cite{bib:Short1}).

\subsection{Defining a Costmap}
Costmaps are typically used to represent a discrete variable distribution (e.g.~traversability) over an area. Costmaps are generated by dividing the environment into grid cells and assigning a single value to each cell. Having one value represent each position within that cell is an approximation and compromises path calculations, especially when the grid cell represents a relatively large area. If the grid cell represents a relatively large area in comparison to the unmanned vehicle (or traveling personnel), then the cell's value used to represent the vehicle's exposure to the environment varies from the actual exposure according to how the costmap variable changes throughout the area of that cell (which for relatively large areas, can change quite drastically). This means that even if the path planner has perfect costmap information, each grid cell will carry some variance in error with respect to the cell's assigned value.

\begin{figure}[htb!] 
\centering
\includegraphics[scale=0.4]{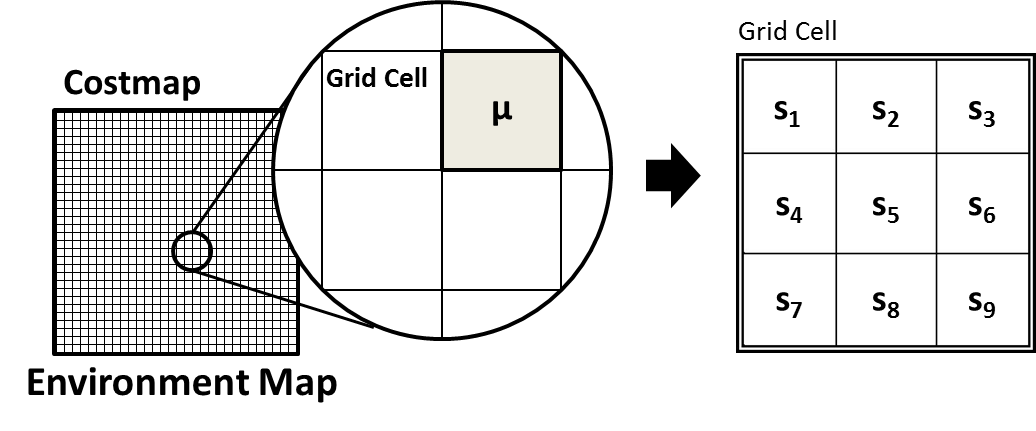}
\caption{Costmap grid cell}
\label{fig:grid_costmap}
\end{figure}

Figure \ref{fig:grid_costmap} illustrates how different positions within a grid cell can experience a variation of the costmap variable even if the true value (denoted as $\mu$) assigned to the cell is known. Figure \ref{fig:grid_costmap} shows a costmap and an individual grid cell. The grid cell is further divided into $N_s$ subcells ($\{s_1,\cdots,s_9\}$) that can each be associated with a unique value representing the distributed costmap variable. For simplification this example uses $N_s=9$, but can be extended to any number of subcells. Given that the values of the subcells are known a mean value $\mu$ and variance $\sigma ^2$ are calculated with equations (\ref{eq:mu}) and (\ref{eq:sig}), respectively.
\begin{align}
\label{eq:mu} \mu &= \frac{1}{N_s}\sum_{i = 1}^{N_s}s_i \\ 
\label{eq:sig} \sigma ^2 &=\frac{1}{N_s}\sum_{i = 1}^{N_s}(s_i - \mu)^2
\end{align}

Generally, costmaps only utilize the mean value by assigning it as the grid cell's costmap value. This work proposes using a two-value costmap by also assigning a variance to each grid cell. This additional information is used to give a measure of uncertainty to routes calculated by the path planner. Doing so allows operators to distinguish routes that have been guaranteed to be safe versus routes that were calculated to be safe but have a high level of uncertainty. This approach is motivated by the work in \cite{bib:RiskLit3}, which uses gaussian process techniques to generate probabilistic costmaps representing disributions over possible terrain costs.

In this work the costmap is divided into an $N_c \times N_c$ cell grid, where $N_c$ represents the number of cells along each dimension of the costmap. Each cell is further divided into $N_s$ subcells. Let $\mathcal{X}^{(N_c \times N_c)} \subset \mathcal{R}^2$ represent the set of subcell information. The subcell information in the $n^{th}$ row and $m^{th}$ column of $\mathcal{X}$ is represented as $X_n^m \in \mathbb{R}^{(N_s \times 1)}$. Let the mean and variance information associated with the subcell information $\mathcal{X}$ be represented as $M\in \mathbb{R}^{(N_c \times N_c)}$ and $V \in \mathbb{R}^{(N_c \times N_c)}$, respectively. Each element $M_{n,m}\in M$ and $V_{n,m}\in V$ reflects the mean and variance values of the subcell information in $X_n^m$.

\section{Method}
\begin{figure}[htb!]
\centering
\includegraphics[scale=0.45]{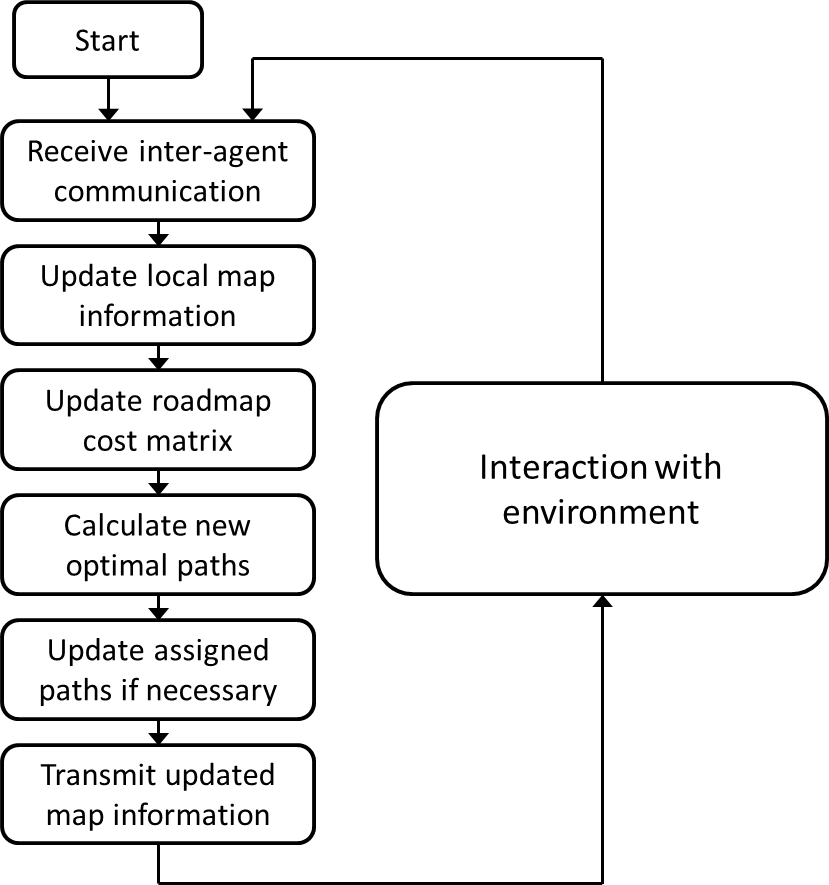}
\caption{Path planning flowchart}
\label{fig:flowchart}
\end{figure}

This work proposes a process for obtaining costmap data and developing the path planning algorithms that use roadmap cost information. This process is described with the flow chart shown in Figure \ref{fig:flowchart}. The proceeding subsections describe our approach in detail.

\subsection{Updating Costmap Information}
\begin{figure}[htb!]
\centering
\includegraphics[scale=0.4]{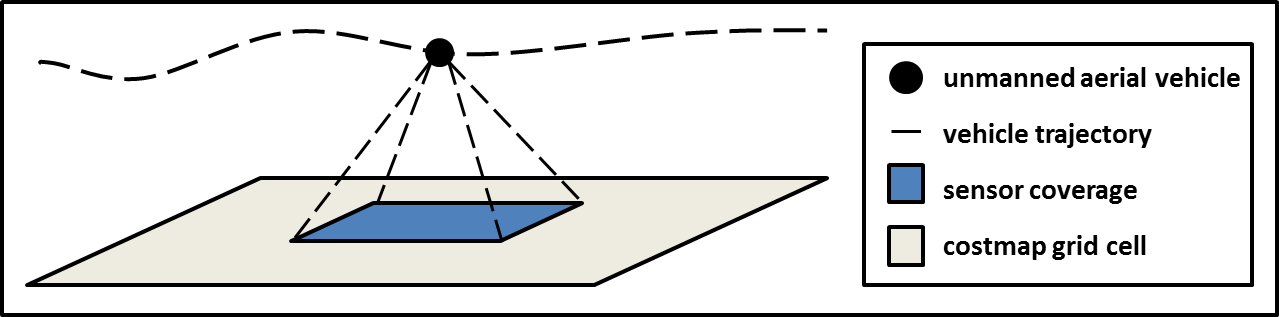}
\caption{Vehicle coverage in cell}
\label{fig:cell_coverage}
\end{figure}

Vehicles that explore unknown areas of the environment update local map information. As the vehicle traverses the grid cell, it collects an amount of information about the cell area. The amount of information collected is determined by the vehicle's sensor coverage and trajectory through the grid cell. If the grid cell represents a relatively large area in comparison to the unmanned vehicle only a portion of the grid cell can be surveyed at any one time. For instance, Figure \ref{fig:cell_coverage} shows a single grid cell, a vehicle, its trajectory over the grid cell, and the vehicle's limited sensor coverage of the cell. An unmanned vehicle (ground vehicle or low flying aerial vehicle) can only survey a fraction of the area defined by the grid cell. 

\begin{figure}[htb!]
\centering
\includegraphics[scale=0.5]{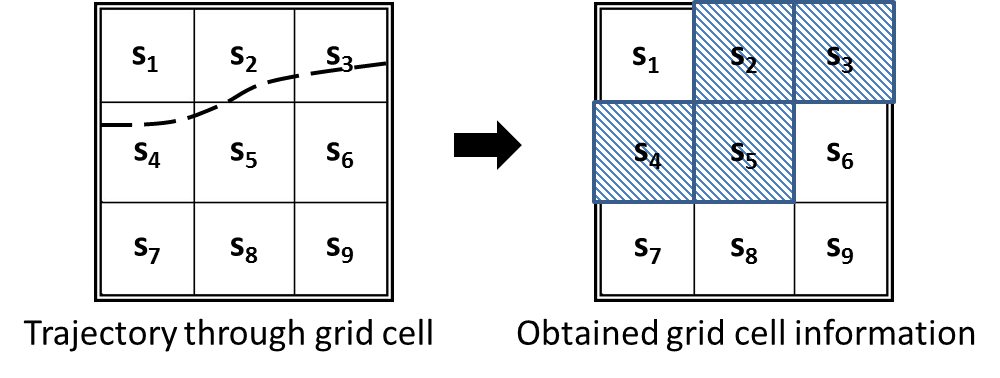}
\caption{Approximation of grid cell coverage}
\label{fig:cell_approx}
\end{figure}

The ratio of `coverage' area to grid cell area depends on the dimensions of the grid cell and limitations of the vehicle's sensor.  To approximate this relationship we have devised some simplifications. Assume that the ratio of coverage to cell area is $1:N_s$ and that the grid cells are divided into subcells according to this coverage ratio. Figure \ref{fig:cell_approx} shows a divided cell given $N_s=9$. The vehicle is assumed to only provide coverage information for the subcell in which it has resided. With this simplification, a trajectory through the grid cell shown in Figure \ref{fig:cell_approx} results in obtained measurement readings for the set of subcells $\{ s_2, s_3, s_4, s_5\}$.

As correct information for any subcell $s_i$ is obtained the previous estimate of the subcell $\hat{s_i}$ is overwritten and the mean and variance information of the grid cell is updated. The mean and variance are updated with the following equations
\begin{align}
\label{eq:update_mu} (\mu)^+ &= (\mu)^- + \frac{(s_k^+ - s_k^-)}{N_s}\\
\label{eq:update_sig}(\sigma^2)^+ &= (\sigma^2)^-  +\frac{(s_k^+ - \mu^+)^2-(s_k^- - \mu^-)^2}{N_s}
\end{align}
where $(\mu)^+$ and $(\sigma^2)^+$ are the updated estimates of the mean and variance, respectively, and $(\mu)^-$ and $(\sigma^2)^-$ are the previous estimates. The obtained measurement of a subcell is denoted as $s_k^+$ and the previous estimate (the \textit{a priori} value) of that subcell is denoted as $s_k^-$. It is assumed there is no noise in obtaining subcell measurements.

Using this update scheme, the estimated cell mean will converge to the true value ($\hat{\mu} \rightarrow \mu$) as the number of unique subcells surveyed reaches $N_s$. In addition, if the true value of the cell mean has already been obtained, the estimated cell variance will converge to the true value ($\hat{\sigma}^2 \rightarrow \sigma^2$) as the number of unique subcells surveyed reaches $N_s$.

\subsection{Calculating Roadmap Scores}
The path planner provides a path as a list of waypoints using roadmap information matrices $G$ and $C$. The calculation of the path relies heavily on how the cost matrix $C$ is defined. This work calculates cost matrix values as a combination of both roadmap and costmap information. The costmap information may be treated as a distribution of terrain traversability, in which case the path planner is designed to: i) reduce the length of the path, ii) reduce the uncertainty of the path, and iii) increase the ease of travel along the path.

\begin{figure}[htb!]
\centering
\includegraphics[scale=0.35]{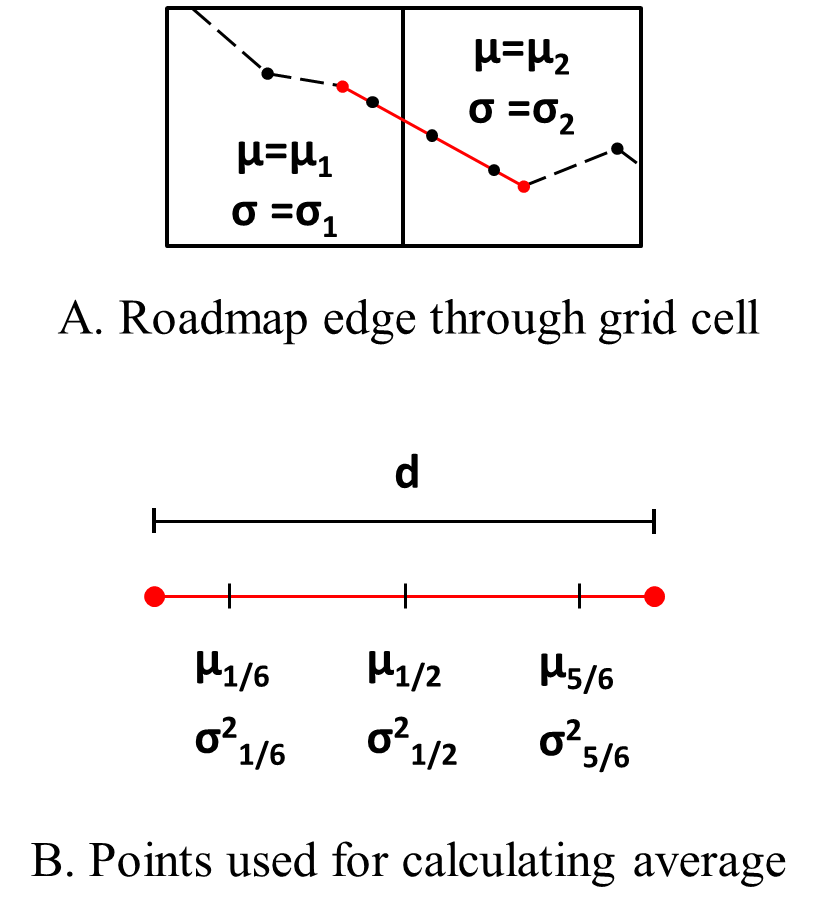}
\caption{Approximation of roadmap edge}
\label{fig:edge_approx}
\end{figure}

Each edge defined in $G$ is scored according to its trajectory through the environment. The edge's score is recorded in $C$. The score is calculated to be proportional to the total length of the edge, and reflects the traversability experienced along the edge. Instead of integrating the length of the edge with respect to these variables, an approximation is used that averages the traversability and its grid cell variance using three points along each edge. Specifically, the points along the edge that are 1/6, 1/2, and 5/6 of the edge length. Figure \ref{fig:edge_approx} illustrates where these points are along some arbitrary edge. 

Figure \ref{fig:edge_approx}-A shows a trajectory through two grid cells, with the red line denoting the edge of interest. This edge is inspected closer in Figure \ref{fig:edge_approx}-B, where three points are identified. With respect to Figure \ref{fig:edge_approx} these points are associated with the $\mu$ values $\mu_{1/6} = \mu_1$, $\mu_{1/2} = \mu_2$, and $\mu_{5/6}=\mu_2$. The values for $\sigma^2$ are assigned similarly. Using these points average values of $\mu$ and $\sigma^2$ used to calculate the edge score are defined as
\begin{align}
\mu_{avg} &=\frac{1}{3}(\mu_{1/6}+\mu_{1/2}+\mu_{5/6}) \\
\sigma^2_{avg} &=\frac{1}{3}(\sigma^2_{1/6}+\sigma^2_{1/2}+\sigma^2_{5/6}) 
\end{align}
where the fraction in the subscript represents its corresponding point along the edge. The edge score $c$ is then defined as

\begin{equation}
c = d(k_1+k_2\mu_{avg}+k_3\sigma^2_{avg})
\label{eq:scoring}
\end{equation}
where $d$ is the total length of the edge and $k_1$,$k_2$, and $k_3$ are weighted gains.

\subsection{Selecting Dissimilar Routes}
In order for the unmanned system to obtain adequate knowledge, the path planner must provide a number of routes that explore unique areas of the map. Each time local costmap information is updated the path planner re-calculates the set of routes using the new vehicle locations as the start points. 

\begin{figure}[htb!]
\centering
\includegraphics[scale=0.35]{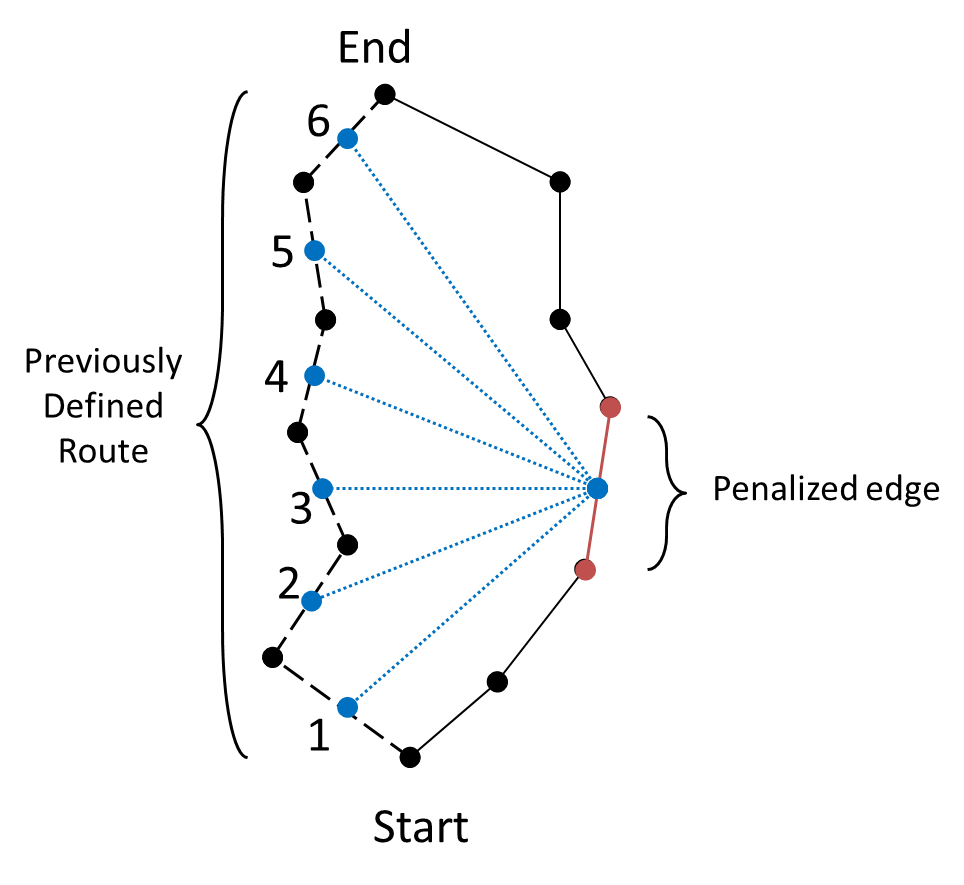}
\caption{Edge penalized due to defined route}
\label{fig:edge_pen}
\end{figure}

The path planner calculates a set of routes in iterations. In each iteration the planner penalizes the roadmap cost matrix $C$ and calculates a single route. In the first iteration a route is calculated using the original roadmap information $G$ and $C$ (i.e.~no penalty). In each other iteration, the cost matrix $C$ is penalized before calculating the route. The cost matrix is penalized by increasing edge scores based on the edge's distance from already defined routes. If the edges are close to any route previously defined then they are heavily penalized, if the edges are far from any route previously defined then they are lightly penalized. Penalizing the cost matrix $C$ in such a manner encourages the planner to find a set of dissimilar routes through the environment. Figure \ref{fig:edge_pen} illustrates how an individual edge is penalized due to its distance from a defined route.

Let $e_i$ be any edge in a defined route. Every edge of the cost matrix $C$ is penalized by a measure of distance between itself and each edge $e_i$ in the defined route (see Figure \ref{fig:edge_pen}). The distance is taken with respect to the midpoints of each edge. Recall that this point coincides with the measurements taken to calculate the roadmap scores. Let $(x_i,y_i)$ be the location of the midpoint for the edge $e_i$, then we define the penalty $f_i(x,y)$ from this edge to a point $(x,y)$ as

\begin{align}
d_i &= \sqrt{(x-x_i)^2+(y-y_i)^2}\\
\label{eq:penal} f_i(x,y) &= \frac{\gamma}{\sqrt{2\pi \bar{\sigma}^2_i}}e^{(-\frac{d_i^2}{2\bar{\sigma}_i})}
\end{align}
The total accumulated penalty due to each edge in the defined route is given as an update of the edge cost.
\begin{equation}
c^+=c^- + \sum_{i=1}^{N_e}f_i(x,y)
\end{equation}
where $c^+$ is the new edge cost after being penalized, and $c^-$ was the previous value. The constant $N_e$ denotes the number of edges in the defined route. The variables $\gamma$ and $\bar{\sigma}^2$ are the control gain and control variance, respectively, that vary the rate at which the edge is penalized due to its distance from previously used edges. Observe that if $\gamma = 0$ no penalty occurs and the alternative routes will be calculated to be equal to the previously defined routes. 

\section{Simulation}

In this section, three experiments are described that provide results regarding the path planning method proposed in this work. While numerous tests may be designed to analyze different aspects of the method (i.e.~computational complexity, stability of the solution, scalability), the experiments presented here are designed to illustrate the fundamental characteristics of the system. The simulations provide results with respect to the multi-vehicle system's capability to: i) use weighted gains to define optimal path conditions, ii) calculate spatially distributed routes with control gain (and control variance), and iii) calculate the most valuable route after exploring the environment. Many of the constraints that may be complex are simplified such that: the time for agents to process information is assumed instantaneous, the dynamic model of the agent is chosen to be linear, the costmap information is 2D and constant, and the system's inter-vehicle communication is guaranteed.

\subsubsection*{Costmap Information}
In every simulation the costmap information is represented by a $20 \times 20$ cell grid. Each $n^{th}$ row and $m^{th}$ column cell is associated with the set of data $X_n^m \in \mathbb{R}^{(9 \times 1)}$. The values for each element of $X_n^m$ are chosen as a random variable from a gaussian distribution with mean and variance values equal to the values of the associated elements in $M$ and $V$ respectively. Each element in $M$ is calculated as a random variable from a uniform distribution in the range $(2,8)$ and $V$ is calculated from a gaussian distribution with zero mean and variance $1$. In addition, each simulation utilizes the parameters listed in Table \ref{table:parameters}.

\begin{table}[]
\centering
\caption{Parameters for each simulation}
\begin{tabular}{|c|c|}
\hline
parameter             & value    \\ \hline \hline
starting position & $[200,1800]$         \\ \hline
target position   & $[1700,600]$     \\ \hline
range of x-axis       & (0:2000)m \\ \hline
range of y-axis       & (0:2000)m \\ \hline
\end{tabular}
\label{table:parameters}
\end{table}

\subsection{Experiment 1}
\begin{figure}[htb!]
\centering
\includegraphics[scale=0.6]{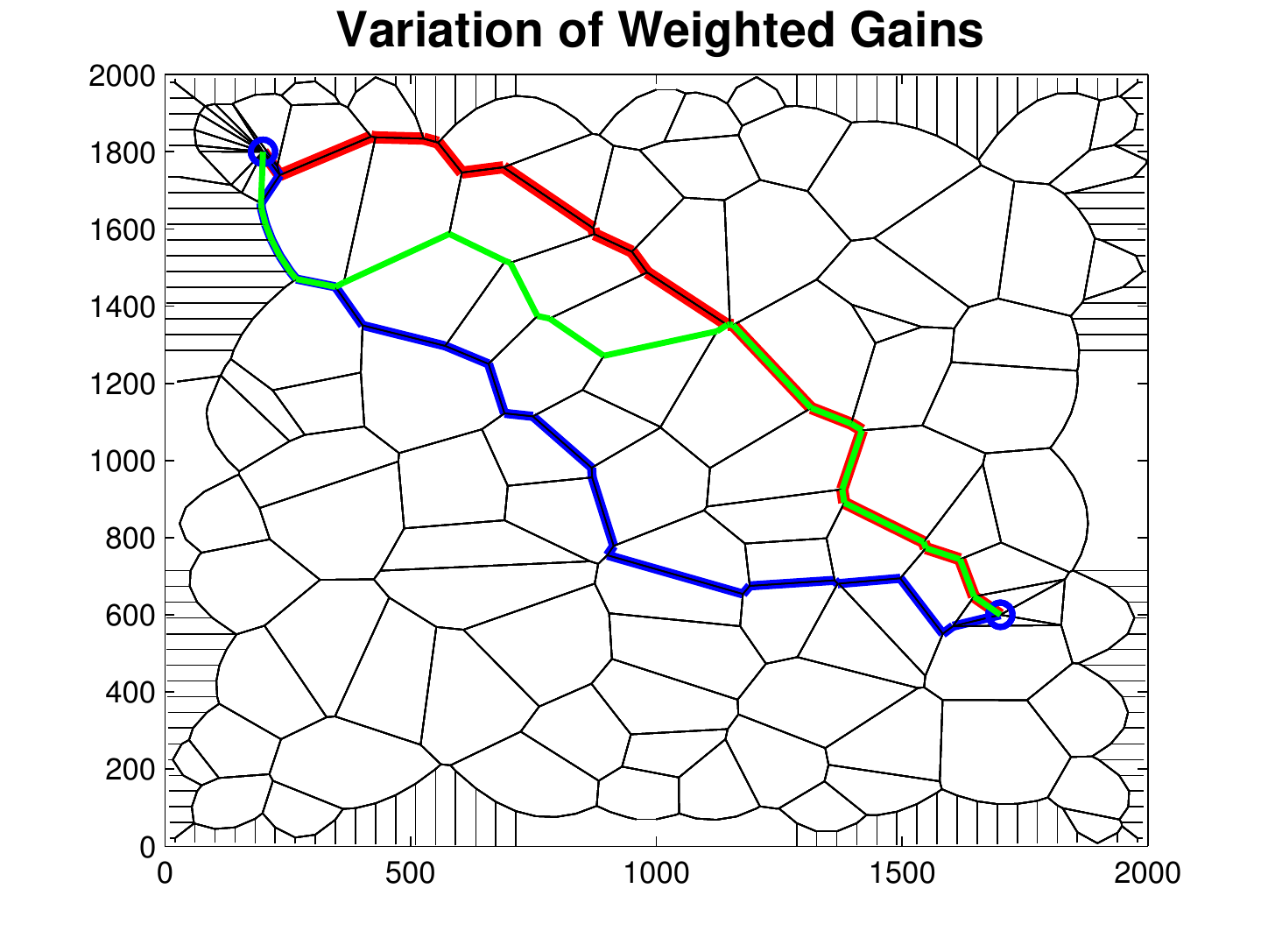}
\caption{Planned paths using different weighted gains}
\label{fig:vary_opt}
\end{figure}

In this experiment a single vehicle plans three paths from a starting point to a target location with perfect costmap information. The trajectory of the paths through the environment is determined by the weighted gains used to score the roadmap information. Each path will be generated using a unique set of weighted gains. Depending on the values of the weighted gains the planner can prioritize the trajectory with respect to distance, terrain traversability, or variation of the traversed areas (see equation \eqref{eq:scoring}). Figure \ref{fig:vary_opt} shows the paths generated by using the weighted gains as defined in Table \ref{table:weight_gains} with control gains set to $\gamma = 10$ and $\bar{\sigma}^2=0.001$.

\begin{table}[htb!]
\centering
\caption{Exp.~1 Parameteters}
\begin{tabular}{cccc} \hline
\multicolumn{1}{|c|}{Path} & \multicolumn{1}{c|}{$k_1$} &  \multicolumn{1}{c|}{$k_2$} &  \multicolumn{1}{c|}{$k_3$} \\ \hline \hline
\multicolumn{1}{|c|}{Red}    & \multicolumn{1}{c|}{30}    & \multicolumn{1}{c|}{10}  & \multicolumn{1}{c|}{10}            \\ \hline
\multicolumn{1}{|c|}{Blue}    & \multicolumn{1}{c|}{0}   & \multicolumn{1}{c|}{30}  & \multicolumn{1}{c|}{0}          \\ \hline
\multicolumn{1}{|c|}{Green}    & \multicolumn{1}{c|}{0}    & \multicolumn{1}{c|}{2}  & \multicolumn{1}{c|}{5}        \\ \hline
\end{tabular}
\label{table:weight_gains}
\end{table}

The path in Figure \ref{fig:vary_opt} outlined in red represents a path that is planned solely on the distance of the route. The path in Figure \ref{fig:vary_opt} outlined in blue represents a path that is planned heavily on the traversability of the route's terrain (although the score is still correlated to the distance). The path Figure \ref{fig:vary_opt} outlined in green represents the lowest terrain variation, meaning that the terrain experienced by traversing the route has a higher consistency with the costmap information used to generate the route.

\subsection{Experiment 2}
\begin{figure}[htb!]
\centering
\includegraphics[scale=0.6]{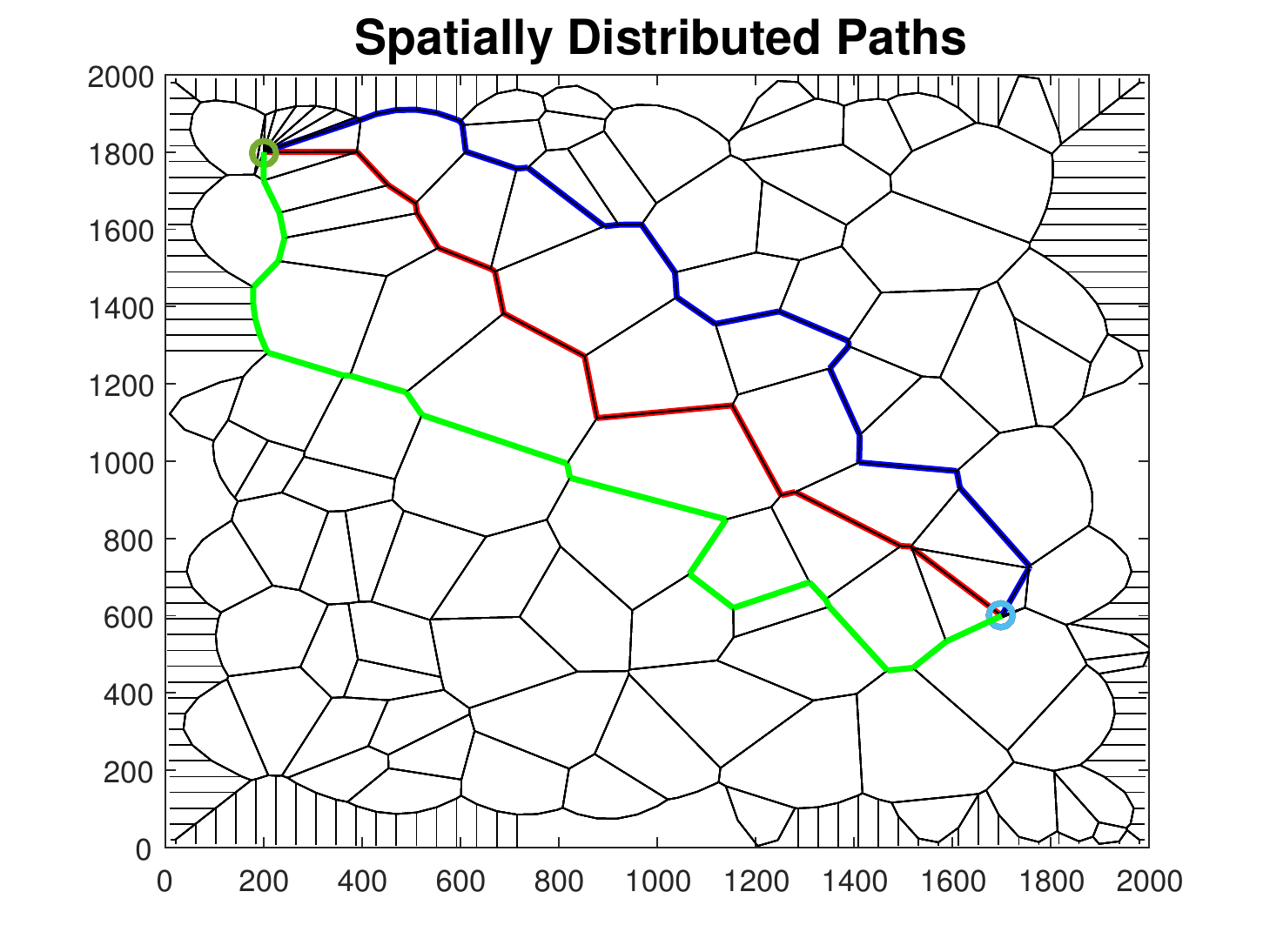}
\caption{Planned paths with low spatial distribution}
\label{fig:low_spatial}
\end{figure}

\begin{figure}[htb!]
\centering
\includegraphics[scale=0.6]{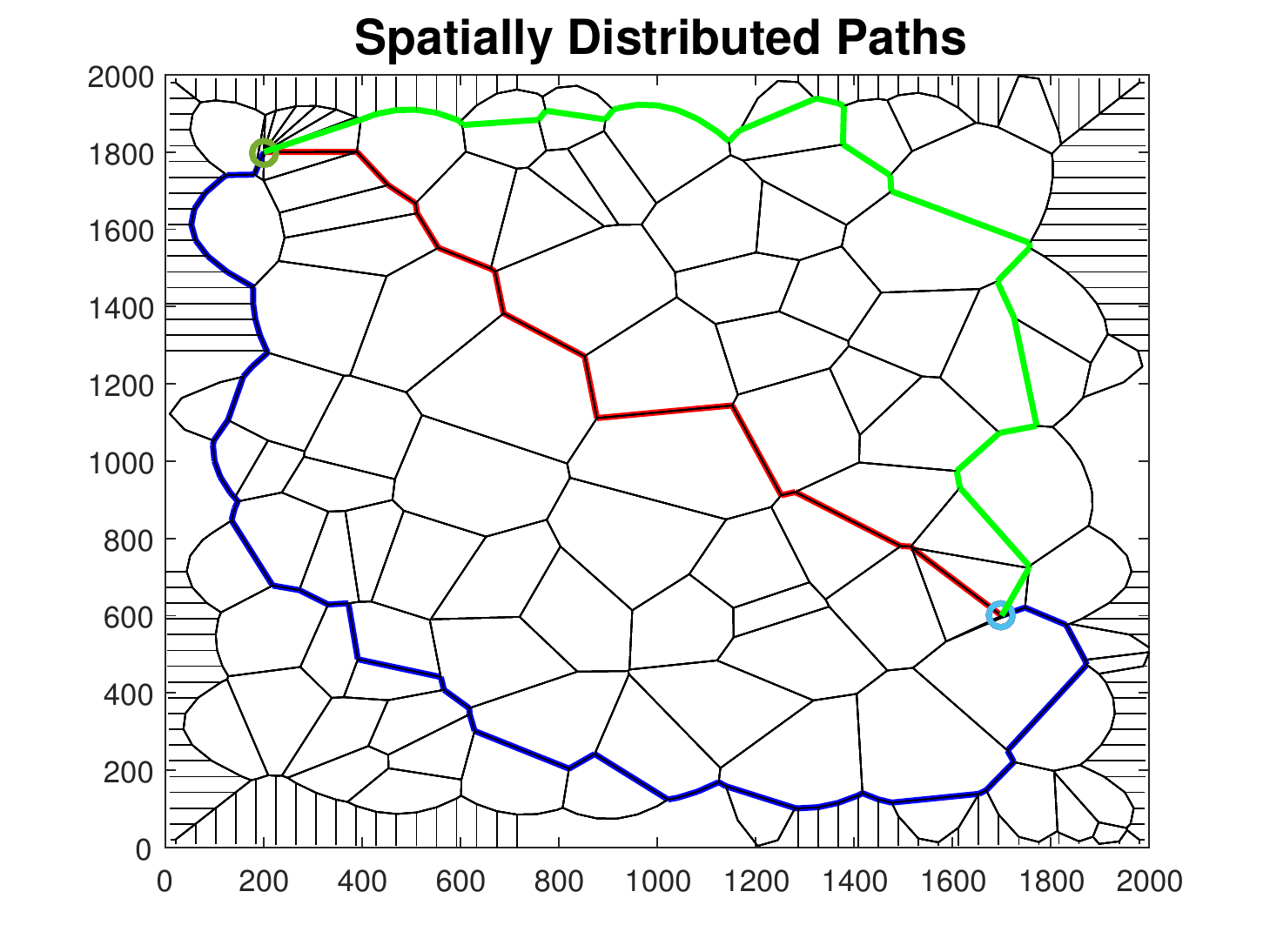}
\caption{Planned paths with high spatial distribution}
\label{fig:high_spatial}
\end{figure}

In this experiment vehicles $A_1$, $A_2$, and $A_3$ cooperatively plan three spatially distributed paths from a starting point to a target location with perfect costmap information. Vehicles plan their paths in sequence, such that each assigned route is generated by feeding penalized roadmap score information to the planner. The score information is penalized according to the vehicle sequence as follows: i) vehicle $A_1$ does not penalize score information, ii) vehicle $A_2$ penalizes score information with respect to the route assigned to $A_1$, and iii) vehicle $A_3$ penalizes the score information with respect to the route assigned to $A_1$ and $A_2$. The spatial distribution (SD) of the paths over the environment is determined by the control gain and control variance. The control gain and control variance are used to penalize the roadmap scores according to the paths assigned to the system (see equation \eqref{eq:penal}). Figure \ref{fig:low_spatial} and \ref{fig:high_spatial} shows the paths generated for $A_1$ (red path), $A_2$ (blue path), and $A_3$ (green path) by using the control gains and weighted gains as defined in Table \ref{table:weight_gains}. Figure \ref{fig:low_spatial} represents the `Low SD' path, where routes are generated with close proximity to each other. Figure \ref{fig:high_spatial} represents the `High SD' path, where routes are generated with far proximity from each other. 

\begin{table}[htb!]
\centering
\caption{Exp.~2 Parameteters}
\begin{tabular}{cccc} \hline
\multicolumn{1}{|c|}{Paths} & \multicolumn{1}{c|}{$\gamma$} &  \multicolumn{1}{c|}{$\bar{\sigma}^2$} &  \multicolumn{1}{c|}{$[k_1~k_2~k_3]$} \\ \hline \hline
\multicolumn{1}{|c|}{Low SD}    & \multicolumn{1}{c|}{100}    & \multicolumn{1}{c|}{0.0003}  & \multicolumn{1}{c|}{[0.6~0.3~0.1]} \\ \hline
\multicolumn{1}{|c|}{High SD}    & \multicolumn{1}{c|}{500}   & \multicolumn{1}{c|}{0.00003} & \multicolumn{1}{c|}{[0.6~0.3~0.1]} \\ \hline
\end{tabular}
\label{table:pen_gains}
\end{table}

\subsection{Experiment 3}
\begin{figure}[htb!]
\centering
\includegraphics[scale=0.45]{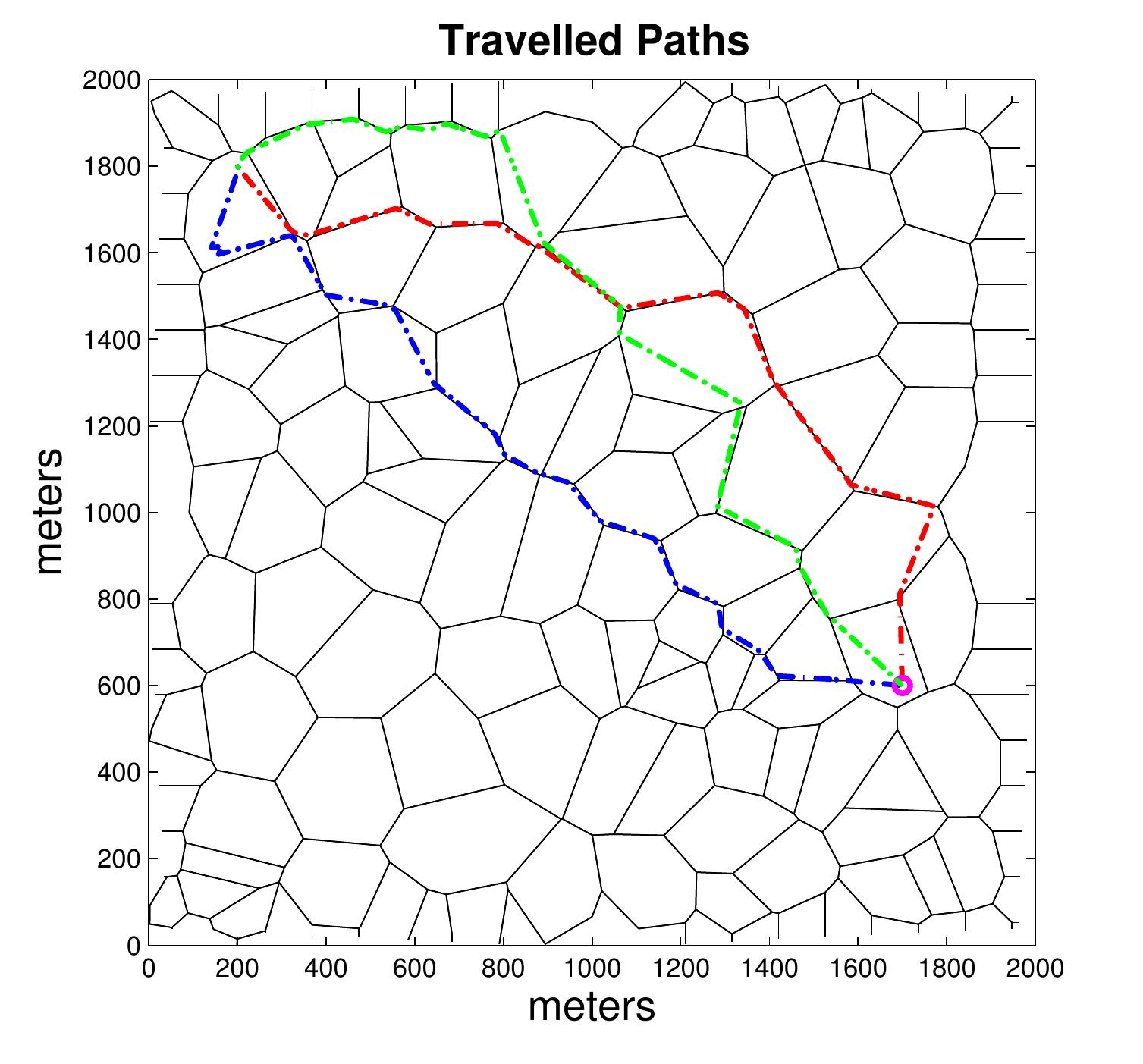}
\caption{Paths travelled by agents}
\label{fig:exp3_trav}
\end{figure}

\begin{figure}[htb!]
\centering
\includegraphics[scale=0.6]{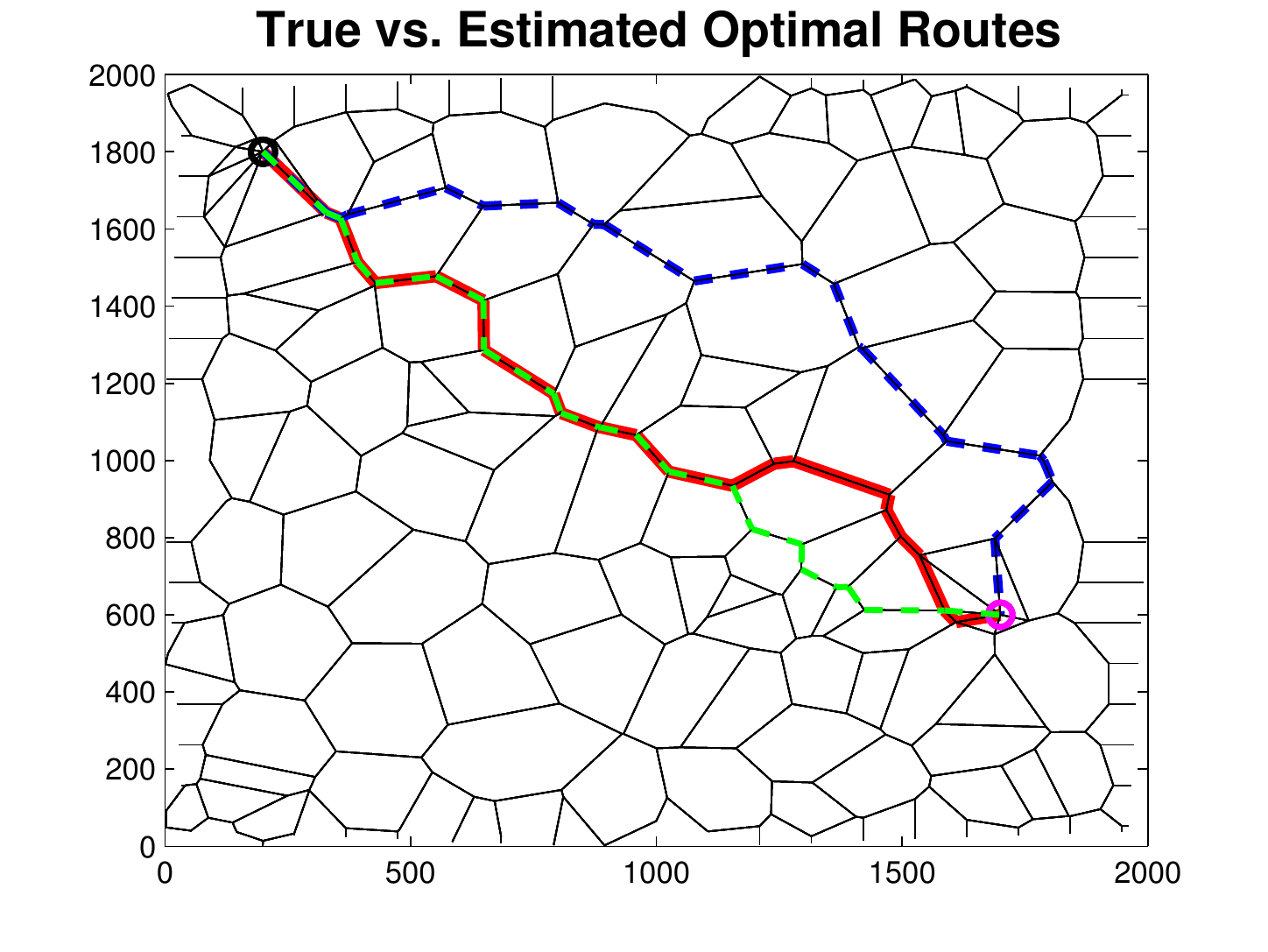}
\caption{True vs estimated optimum paths}
\label{fig:exp3_opt}
\end{figure}

In this experiment vehicles $A_1$, $A_2$, and $A_3$ cooperatively plan three spatially distributed paths from a starting point to a target location using estimated costmap information. As vehicles explore the environment they update the estimated costmap information (see equations \eqref{eq:update_mu} and \eqref{eq:update_sig}). After reaching the target location, the updated costmap information is used to calculate the most valuable route from starting point to the target location. Figure \ref{fig:exp3_trav} shows the trajectories of agent $A_1$ in red, $A_2$ in blue, and $A_3$ in green. Figure \ref{fig:exp3_opt} compares the trajectories between three optimal routes. Two of the routes are calculated using the \textit{a priori} and updated estimate of the costmap information. The third route is calculated using the true costmap information. The control gains and weighted gains used for providing these results are defined in Table \ref{table:exp3}.

\begin{table}[htb!]
\centering
\caption{Exp.~3 Parameteters}
\begin{tabular}{cccc} \hline
\multicolumn{1}{|c|}{$\gamma$} &  \multicolumn{1}{c|}{$\bar{\sigma}^2$} &  \multicolumn{1}{c|}{$[k_1~k_2~k_3]$} \\ \hline \hline
\multicolumn{1}{|c|}{10}    & \multicolumn{1}{c|}{0.0001}  & \multicolumn{1}{c|}{[0~1~0]} \\ \hline
\end{tabular}
\label{table:exp3}
\end{table}

As seen in Figures \ref{fig:exp3_trav} the paths assigned to the vehicles change throughout the mission. The vehicles continuously update data and plan new paths. Figure \ref{fig:exp3_opt} shows three routes. The solid red route is the true optimal route. The blue dashed route is the route calculated using the \textit{a priori} information. The green dashed route is the route calculated using the updated estimated information. Observe that the updated estimated information enables the planner to provide a more optimal route then what would have been provided by only using the \textit{a priori} data. Also, notice from the value of the weighted gains that the routes are evaluated by prioritizing the costmap value (in this case traversability).

\section{Conclusion}
Planning a good quality path in dynamic environments requires exact map information and a solid definition of what the optimal route conditions are. The path planning method proposed assumes that estimated \textit{a priori} map information is available and updated through exploration of the environment. The path planning method uses weights to prioritize the optimal route conditions, and a map update scheme to address sensor coverage limitations. To increase coverage the planner generates spatially diverse routes for the vehicle to travel. The spatial distribution of the paths is varied using control gains in a penalization process. The planner penalizes roadmap information used to generate the vehicle's path based on routes assigned to other vehicles. After traversing their routes, from a starting point to a target location, the system calculates a candidate route to be the best quality route according to the defined optimal conditions and updated map information.

To emphasize the utilization of the proposed method simulations are provided that observe how varying different parameters effects the performance. Simulations are conducted to test the ability of the path planner by: i) varying the weighted gains, ii) varying the control gains, and iii) comparing the system's best estimated route with the true optimal route. Results show that the performance of the proposed method relies heavily on how the true and estimated costmap information is defined. Due to the vehicle's partial coverage of the costmap cells, the update scheme does not allow the estimated costmap values to converge to their true values. This can cause the candidate route to be incorrectly identified. Further investigation is needed to identify the relationship between parameter values and how the correlation between different costmap information can effect path planning performance.

\bibliographystyle{IEEEtran}
\bibliography{refs}

\end{document}